\documentstyle[12pt,amssymbols]{article}
\def\journal{\topmargin .3in	\oddsidemargin .5in
	\headheight 0pt	\headsep 0pt
	\textwidth 5.625in 
	\textheight 8.25in 
	\marginparwidth 1.5in
	\parindent 2em
	\parskip .5ex plus .1ex		\jot = 1.5ex}
\journal

\catcode`\@=11
\def\marginnote#1{}
\newcount\hour
\newcount\minute
\newtoks\amorpm
\hour=\time\divide\hour by60
\minute=\time{\multiply\hour by60 \global\advance\minute by-\hour}
\edef\standardtime{{\ifnum\hour<12 \global\amorpm={am}%
	\else\global\amorpm={pm}\advance\hour by-12 \fi
	\ifnum\hour=0 \hour=12 \fi
	\number\hour:\ifnum\minute<10 0\fi\number\minute\the\amorpm}}
\edef\militarytime{\number\hour:\ifnum\minute<10 0\fi\number\minute}
\def\draftlabel#1{{\@bsphack\if@filesw {\let\thepage\relax
   \xdef\@gtempa{\write\@auxout{\string
      \newlabel{#1}{{\@currentlabel}{\thepage}}}}}\@gtempa
   \if@nobreak \ifvmode\nobreak\fi\fi\fi\@esphack}
	\gdef\@eqnlabel{#1}}
\def\@eqnlabel{}
\def\@vacuum{}
\def\draftmarginnote#1{\marginpar{\raggedright\scriptsize\tt#1}}
\def\draft{\oddsidemargin -.5truein
	\def\@oddfoot{\sl preliminary draft \hfil
	\rm\thepage\hfil\sl\today\quad\militarytime}
	\let\@evenfoot\@oddfoot	\overfullrule 3pt
	\let\label=\draftlabel
	\let\marginnote=\draftmarginnote
   \def\@eqnnum{(\theequation)\rlap{\kern\marginparsep\tt\@eqnlabel}%
\global\let\@eqnlabel\@vacuum}  }
\def\preprint{\twocolumn\sloppy\flushbottom\parindent 2em
	\leftmargini 2em\leftmarginv .5em\leftmarginvi .5em
	\oddsidemargin -.5in	\evensidemargin -.5in
	\columnsep .4in	\footheight 0pt
	\textwidth 10in	\topmargin  -.4in
	\headheight 12pt \topskip .4in
	\textheight 7.1in \footskip 0pt
	\def\@oddhead{\thepage\hfil\addtocounter{page}{1}\thepage}
	\let\@evenhead\@oddhead	\def\@oddfoot{}	\def\@evenfoot{} }
\def\numberbysection{\@addtoreset{equation}{section}
	\def\theequation{\thesection.\arabic{equation}}}
\def\underline#1{\relax\ifmmode\@@underline#1\else
	$\@@underline{\hbox{#1}}$\relax\fi}
\def\titlepage{\@restonecolfalse\if@twocolumn\@restonecoltrue\onecolumn
     \else \newpage \fi \thispagestyle{empty}\c@page\z@
	\def\thefootnote{\fnsymbol{footnote}} }
\def\endtitlepage{\if@restonecol\twocolumn \else \newpage \fi
	\def\thefootnote{\arabic{footnote}}
	\setcounter{footnote}{0}}  
\catcode`@=12
\relax
\def\figcap{\section*{Figure Captions\markboth
	{FIGURECAPTIONS}{FIGURECAPTIONS}}\list
	{Figure \arabic{enumi}:\hfill}{\settowidth\labelwidth{Figure 999:}
	\leftmargin\labelwidth
	\advance\leftmargin\labelsep\usecounter{enumi}}}
 \relax
\def\tablecap{\section*{Table Captions\markboth
	{TABLECAPTIONS}{TABLECAPTIONS}}\list
	{Table \arabic{enumi}:\hfill}{\settowidth\labelwidth{Table 999:}
	\leftmargin\labelwidth
	\advance\leftmargin\labelsep\usecounter{enumi}}}
 \relax
\def\reflist{\section*{References\markboth
	{REFLIST}{REFLIST}}\list
	{[\arabic{enumi}]\hfill}{\settowidth\labelwidth{[999]}
	\leftmargin\labelwidth
	\advance\leftmargin\labelsep\usecounter{enumi}}}
 \relax
\makeatletter
\newcounter{pubctr}
\def\publist{\@ifnextchar[{\@publist}{\@@publist}}
\def\@publist[#1]{\list
	{[\arabic{pubctr}]\hfill}{\settowidth\labelwidth{[999]}
	\leftmargin\labelwidth
	\advance\leftmargin\labelsep
	\@nmbrlisttrue\def\@listctr{pubctr}
	\setcounter{pubctr}{#1}\addtocounter{pubctr}{-1}}}
\def\@@publist{\list
	{[\arabic{pubctr}]\hfill}{\settowidth\labelwidth{[999]}
	\leftmargin\labelwidth
	\advance\leftmargin\labelsep
	\@nmbrlisttrue\def\@listctr{pubctr}}}
 \relax
\makeatother
\catcode`\@=11
\def\section{\@startsection {section}{1}{0pt}{-3.5ex plus -1ex minus
 -.2ex}{2.3ex plus .2ex}{\raggedright\large\bf}}
\catcode`\@=12
\newskip\humongous \humongous=0pt plus 1000pt minus 1000pt
\def\caja{\mathsurround=0pt}

\newif\ifdtup
\def\panorama{\global\dtuptrue \openup1\jot \caja
	\everycr{\noalign{\ifdtup \global\dtupfalse
	\vskip-\lineskiplimit \vskip\normallineskiplimit
	\else \penalty\interdisplaylinepenalty \fi}}}
\def\eqalignno#1{\panorama \tabskip=\humongous
	\halign to\displaywidth{\hfil$\displaystyle{##}$
	\tabskip=0pt&$\displaystyle{{}##}$\hfil
	\tabskip=\humongous&\llap{$##$}\tabskip=0pt
	\crcr#1\crcr}}

\def\oldreffmt#1{\rlap{[#1]} \hbox to 2\parindent{}}

\def\figfmt#1{\rlap{Figure {#1}} \hbox to 1in{}}

\def\beq{\begin{equation}}
\def\eeq{\end{equation}}

\def\bea{\begin{eqnarray}}

\def\eea{\end{eqnarray}}
\catcode`\@=11
\def\eqnarray{\stepcounter{equation}\let\@currentlabel=\theequation
\global\@eqnswtrue
\global\@eqcnt\z@\tabskip\@centering\let\\=\@eqncr
\gdef\@@fix{}\def\eqno##1{\gdef\@@fix{##1}}%
$$\halign to \displaywidth\bgroup\@eqnsel\hskip\@centering
  $\displaystyle\tabskip\z@{##}$&\global\@eqcnt\@ne
  \hskip 2\arraycolsep \hfil${##}$\hfil
  &\global\@eqcnt\tw@ \hskip 2\arraycolsep $\displaystyle\tabskip\z@{##}$\hfil
   \tabskip\@centering&\llap{##}\tabskip\z@\cr}

\def\@@eqncr{\let\@tempa\relax
    \ifcase\@eqcnt \def\@tempa{& & &}\or \def\@tempa{& &}
      \else \def\@tempa{&}\fi
     \@tempa \if@eqnsw\@eqnnum\stepcounter{equation}\else\@@fix\gdef\@@fix{}\fi
     \global\@eqnswtrue\global\@eqcnt\z@\cr}

\catcode`\@=12
\font\tenbifull=cmmib10 
\font\tenbimed=cmmib10 scaled 800
\font\tenbismall=cmmib10 scaled 666
\relax

\def\thefootnote{\fnsymbol{footnote}}
\begin{document}
\begin{titlepage}
\begin{center}
\today     \hfill    LBL-35842 \\
          \hfill    UCB-PTH-94/18 \\

\vskip .5in

{\large \bf The Conformal Points Of The Generalized Thirring Model}
\footnote{This work was supported in part by the Director, Office of
Energy Research, Office of High Energy and Nuclear Physics, Division of
High Energy Physics of the U.S. Department of Energy under Contract
DE-AC03-76SF00098 and in part by the National Science Foundation under
grant PHY-90-21139.}

\vskip .5in

Korkut Bardakci\\
\vskip .5in

{\em Theoretical Physics Group\\
    Lawrence Berkeley Laboratory\\
      University of California\\
    Berkeley, California 94720}
\end{center}

\vskip .5in

\begin{abstract}

The conformal fixed points of the generalized Thirring model are
investigated with the help of bosonization, the large N limit and the
operator product expansion. Necessary conditions on the coupling constants
for conformal invariance are derived.
\end{abstract}
\end{titlepage}
\renewcommand{\thepage}{\roman{page}}
\setcounter{page}{2}
\mbox{ }

\vskip 1in

\begin{center}
{\bf Disclaimer}
\end{center}

\vskip .2in

\begin{scriptsize}
\begin{quotation}
This document was prepared as an account of work sponsored by the United
States Government. While this document is believed to contain correct
 information, neither the United States Government nor any agency
thereof, nor The Regents of the University of California, nor any of their
employees, makes any warranty, express or implied, or assumes any legal
liability or responsibility for the accuracy, completeness, or usefulness
of any information, apparatus, product, or process disclosed, or represents
that its use would not infringe privately owned rights.  Reference herein
to any specific commercial products process, or service by its trade name,
trademark, manufacturer, or otherwise, does not necessarily constitute or
imply its endorsement, recommendation, or favoring by the United States
Government or any agency thereof, or The Regents of the University of
California.  The views and opinions of authors expressed herein do not
necessarily state or reflect those of the United States Government or any
agency thereof or The Regents of the University of California and shall
not be used for advertising or product endorsement purposes.
\end{quotation}
\end{scriptsize}

\vskip 2in

\begin{center}
\begin{small}
{\it Lawrence Berkeley Laboratory is an equal opportunity employer.}
\end{small}
\end{center}

\newpage
\renewcommand{\thepage}{\arabic{page}}
\setcounter{page}{1}
\noindent {\bf 1.Introduction}
\vskip 9pt

At present, the study of two dimensional conformal field theories is an
active branch of mathematical physics. These theories, as well as being of
interest on their own [1], play an important role in string compactification
 [2]. Although, unfortunately, there is as yet no complete classification of
two dimensional conformal field theories , many important examples are known
and have been studied over the years. Among these are free fields, the Wess-
Zumino-Witten(WZW) [3] model, the gauged version of WZW[4], and models based
on Calabi-Yau manifolds [5]. All of these models have the standard local
lagrangian description and as a result, they  have been  used in
 various schemes of string
compactification. There are also conformal theories based on Hamiltonians
built out of affine chiral currents (affine Virasoro construction)[6,7],
 which await application in string theory.

Among the natural candidates for conformal theories is the generalized
Thirring model in two dimensions, which in this paper is defined as a
model of several massless
fermions interacting through the most general Lorentz invariant four
fermion couplings. There are several reasons for being interested in this
class of models. One of the simplest types of string compactification
makes use of free fermions, and the most natural renormalizable generalization
of the free Fermion theory consists of adding a four Fermi interaction.
 One can therefore hope to
construct new string theories based on  compactification through some
generalized version of the Thirring model. It is also possible that new
string theories of this type can help us understand QCD better. There is
good evidence that QCD in the large N limit reduces to a theory of non-
interacting strings [8]. It is also clear that these strings cannot be
 described
by the well known standard string theory, not at least without important
modifications. The present author has proposed as a candidate [9] for the QCD
string theory the following world sheet action which is the sum of three
terms: The first term is the standard free field action for the target space
coordinates. The second term is a generalized Thirring model built out of
the fermionic partners of the bosonic fields, and the third term is a
Yukawa interaction coupling the bosons to the fermions. This therefore
provides another motivation for studying the generalized Thirring model
in the context of string theory.

It is well known that the first condition for constructing a satisfactory
 string model is to start with a local field theory on the world sheet that
is conformally invariant [1]. One is therefore naturally led to investigate
the conditions that have to be imposed on the coupling constants in order to
have a conformally invariant generalized Thirring model. In a pioneering
paper, Dashen and Frishman [10] showed that in the special case of a four
Fermi interaction invariant under a  Lie group (non-abelian Thirring model),
the requirement of conformal invariance results in the quantization of the
coupling constant. Since then,  there has been more work on both
the original model [11, 12] of  Dashen and Frishman and also on its
 generalizations [13, 14]. Although the original results of Dashen and
Frishman are confirmed, it is clear that a full understanding of the
 conformal invariance of the
generalized Thirring model still remains as an open problem.

This paper is another attempt to find the conformal fixed points the
Thirring model. As in most of the previous work, we start by bosonizing
the model, and as a preliminary step to quantization, we work out the
Poisson bracket (PB) algebra in the light cone coordinates. Next, we try to
 quantize again in the light cone frame, by converting the
 PB's into the operator product expansion (OPE). This is done by translating
singular terms in the coordinate differences in the PB's into their analogues
in the OPE's(eq.(4.4a)). We  are, however, only
able to do this term by term in a large N expansion, and here we compute
only the first two terms, although higher order terms are in principle
 calculable.
As a result, we have, at least to the given order in N, a fully consistent
algebra which is a natural generalization of the affine algebra, and which
may be of interest on its own.
Conformal invariance is then imposed by constructing the stress tensor and
requiring that it satisfy the Virasoro algebra. This results in algebraic
conditions on the coupling constants of the Thirring model. Similar
conditions were derived in reference [14] from the $\beta$ function
equations in the one loop approximation [15]. Although, in principle, our
 results should agree with those of [14], we found it difficult to make a
direct comparison. This is due to the difference in the methods used (
Hamiltonian approach in this paper as opposed to the Lagrangian approach
in [14]) and the consequent
 difficulty of comparing the definitions of renormalized coupling
constants.
\vskip 9pt
\noindent {\bf 2. Bosonization of the Generalized Thirring Model}
\vskip 9pt

 It is by now well known how to bosonize general two dimensional
fermionic field theories [4,16], and the case of the generalized Thirring
model was treated in, for example, in reference [13]. For the sake of
completeness and to establish notation, we will present here a brief
treatment. The starting point is the action
$$
S=\int d^{2}x(\bar{\Psi}i\gamma^{\mu}\partial_{\mu}\Psi-\tilde{G}_{ab}^{-1}
\; \bar{\Psi}
\gamma_{\mu}\lambda_{a}\Psi\;\bar{\Psi}\gamma^{\mu}\lambda_{b}\Psi),\eqno(2.1)
$$
where $\tilde{G}_{ab}$ are the coupling constants, and $\Psi$ is a Dirac
 fermion in the fundamental representation of $SU(n)$,(or $U(n)$),
 considered as a flavor group.
The $\lambda's$ are matrices in the adjoint of the same group, they are
 trace orthogonal and satisfy the commutation relations
$$
[\lambda_{a},\lambda_{b}]=f_{abc}\lambda_{c}.\eqno(2.2)
$$
In addition to the flavor group, which is in general broken by the coupling
constant matrix G, there is also a color group $U(N)$, which is an exact
symmetry of the model. The fermions are again in the fundamental of this
group, and the color indices are contracted in  the fermion
bilinears in (2.1)  to form singlets. In what follows, the large N
limit will be helpful in making the model tractable.

We now introduce auxiliary fields $A^{\mu}_{a}$ to rewrite the action as
$$
S=\int d^{2}x(\bar{\Psi}i\gamma^{\mu}\partial_{\mu}\Psi+\bar{\Psi}\gamma^{\mu}
\lambda_{a}\Psi A_{\mu,a}+\frac{1}{4}\tilde{G}_{ab}A^{\mu}_{a}A_{\mu,b}).
\eqno(2.3)
$$
Using the Polyakov-Wiegmann formula [16], the functional integration over the
fermion field can be carried out and the resulting determinant can be
computed. It is convenient to express the resulting action in terms of
unitary matrices $g(x)$ and $h(x)$ defined by
$$
A_{+}=ig^{-1}\partial_{+}g,\; A_{-}=ih^{-1}\partial_{-}h,\eqno (2.4a)
$$
where
$$
\partial_{\pm}=\partial_{0} \mp \partial_{1},\: A_{\pm}=A_{0} \mp A_{1},
\eqno (2.4b)
$$
and the matrix notation
$$
A_{\pm}=\lambda_{a} A_{\pm,a} \eqno (2.4c)
$$
has been used. In terms of the fields g and h, the action takes on the
following form:
$$
S=\int d^{2}x\bar{\Psi} i \gamma^{\mu}\partial_{\mu}\Psi+W(g)+W(h^{-1})
-\frac{N}{2\pi}\int d^{2}x G_{ab}(g^{-1}\partial_{+}g)_{a}(h^{-1}
\partial_{-}h)_{b},\eqno(2.5a)
$$
where W is the WZW action
$$
W(g)=\frac{N}{8\pi}\int d^{2}x Tr(g^{-1}\partial_{\mu}g\,g^{-1}\partial^
{\mu}g)-\frac{N}{12\pi}\int Tr\Bigl((g^{-1}dg)^{3}\Bigr).\eqno(2.5b)
$$
Here and in what follows,  $X_{a}$ stands for
 $Tr(\lambda_{a}X)$. The relation between G and $\tilde{G}$ is
$$
\frac{N}{2\pi}G_{ab}=\frac{N}{4\pi}\delta_{ab}+\frac{1}{4}\tilde{G_{ab}}.
\eqno(2.6)
$$
It should also be understood that, to take care of the Jacobian resulting from
the change of variables from $A_{\mu}$ to g and h, the N that appears in these
 equations should be shifted by the Casimir of the group.

In the next section, we shall need the variation of the action  (2.5), which
is written down below:
$$
\eqalignno{
\delta\!S=\frac{N}{2\pi}\int &Tr\Bigl(\delta\!g g^{-1}\partial_{+}(-\frac{1}{2}
\partial_{-}g g^{-1}+g\lambda_{a}g^{-1}G_{ab}(h^{-1}\partial_{-}h)_{b} \cr
+& \delta\!h h^{-1}\partial_{-}(- \frac{1}{2}\partial_{+}h h^{-1}+h\lambda_{b}
h^{-1}G_{ab}(g^{-1}\partial_{+}g)_{a})\Bigr).&(2.7)\cr}
$$
We also note that the equations of motion yield two chiral currents $J_{\pm}$:
$$
\partial_{+}J_{-}=\partial_{-}J_{+}=0,\eqno(2.8a)
$$
where,
$$
\eqalignno{
J_{+}=& i\frac{N}{4\pi}\Bigl(-\partial_{+}h h^{-1}+2 h\lambda_{b}h^{-1} G_{ab}
(g^{-1}\partial_{+}g)_{a}\Bigr),\cr
J_{-}=& i\frac{N}{4\pi}\Bigl(-\partial_{-}g g^{-1}+2 g\lambda_{a}g^{-1}
G_{ab}(h^{-1}\partial_{-}h)_{b}\Bigr).&(2.8b)\cr}
$$
These  currents are conserved by virtue of invariance of the action under
$$
g\rightarrow u_{-}(x_{-})g,\: h\rightarrow u_{+}(x_{+})h,\eqno(2.9)
$$
where $x_{\pm}=\frac{1}{2}(x_{0}\mp x_{1})$ and $u_{\pm}$ are arbitrary
functions.
\vskip 9pt
\noindent {\bf 3. The Poisson Bracket Algebra}
\vskip 9pt

{}From the variation of the action presented in the last section(eq. 2.7), it
is easy to read off the the basic PB's of the field variables. We shall as
before use light cone variables, treating $x_{+}$ as time and $x_{-}$ as
space. We remind the reader that a first order action of the form
$$
S=\int dt \Bigl(\frac{d\phi^{a}}{dt}A_{a}(\phi)-V(\phi)\Bigr), \eqno(3.1a)
$$
with the variation,
$$
\delta\!S=\int \delta\!\phi^{a}\Bigl(\frac{d\phi^{b}}{dt}K_{ab}(\phi)
-\frac{\partial V}{\partial\phi^{a}}\Bigr),\eqno(3.1b)
$$
has the PB algebra given by
$$
\{\phi^{a},\phi^{b}\}=F^{ab}=(K^{-1})^{ab}.\eqno(3.2b)
$$
In our case, the variables whose PB's are the most convenient to compute
 (3,17) are the tangent space one forms ${\cal M}^{i}_{a}$, defined by
$$
{\cal M}^{1}_{a} \equiv (i g^{-1}\delta\! g)_{a},\:
{\cal M}^{2}_{a}\equiv (i \delta\! h h^{-1})_{a}. \eqno (3.3a)
$$
The corresponding functions $K^{(ij)}_{ab}(x,y)$, $i,j=1,2$, can be read off
from the variation of the action,
$$
\delta\! S=\int dx_{+}\; dx\; dy {\cal M}^{i}_{a}(x) K^{(ij)}_{ab}(x,y)
{\cal M}^{j}_{b}(y),\eqno (3.3b)
$$
and by comparing it with eq.(2.7). In this equation and in what follows,
 to simplify the notation,  the subscript (-) on the space variables
$x_{-}$ and $y_{-}$ is dropped. The PB's are then given by
$$
\{{\cal M}^{i}_{a}(x),{\cal M}^{j}_{b}(y)\}= F^{(ij)}_{ab}(x,y),\eqno (3.4)
$$
where F is the matrix inverse of K, as in eq. (3.2b). The end result is that,
after some straightforward algebra, one obtains the following differential
equations for F:
$$
\eqalignno{
\partial_{x}F^{(11)}_{ab}(x,y)& -  2 G_{kl} m_{l}(x) f_{kac}
 F^{(11)}_{cb}(x,y) -  2 G_{ac} R_{cd}(x)\partial_{x} F^{(21)}_{db}(x,y)\cr
 = \frac{4 \pi}{N} \delta_{ab} & \delta(x-y),\cr
\partial_{x} F^{(12)}_{ab} (x,y)& -  2 G_{kl} m_{l}(x) f_{kac}
F^{(12)}_{cb}(x,y)   - 2 G_{ac}R_{cd}(x) \partial_{x} F^{(22)}_{db}(x,y)
\cr =& 0, \cr
\partial_{x} \Bigl(2 R_{ca}(x)& G_{cd} F^{(11)}_{db}(x,y)  -
F^{(21)}_{ab}(x,y)\Bigr)=0,\cr
\partial_{x}\Bigl(2 R_{ca} (x)& G_{cd} F^{(12)}_{db}(x,y)
- F^{(22)}_{ab}(x,y)\Bigr)= \frac{4 \pi}{N} \delta_{ab} \delta(x-y),
& (3.5a) \cr}
$$
where,
$$
R_{ab}(x)= Tr (\lambda_{a} h^{-1}(x)\lambda_{b}h(x)),\;
m_{l}(x)= Tr(i h^{-1}(x) h'(x)\lambda_{l}).\eqno(3.5b)
$$
These equations can easily be integrated. To write the result in a compact
form, we define,
$$
\eqalignno{
H_{ab}& =\delta_{ab}-4 (G^{2})_{ab}, \cr
(Q_{c})_{ab}& = - 2 G_{c c'} f_{c'ab}+ 4 G_{a a'} G_{b b'} f_{a' b' c},\cr
(B_{c})_{ab}& =( H^{-\frac{1}{2}} Q_{c} H^{-\frac{1}{2}})_{ab},& (3.6)\cr}
$$
and we supress the indices a and b in quantities like
 $ (Q_{c})_{ab}$, $R_{ab}$ and $ F_{ab}$  and use the matrix notation.
 With this convention, we have,
$$
\eqalignno{
F^{(11)}(x,y)& = \frac{2 \pi}{N} R^{T}(x) G H^{-\frac{1}{2}}
P exp \Bigl(\int_{x}^{y} dx' B_{c} m_{c}(x')\Bigr) H^{-\frac{1}{2}}
\; \varepsilon (x-y),\cr
F^{(12)}(x,y)&=\frac{4 \pi}{N} P exp \Bigl(\int_{x}^{y} dx' B_{c}
m_{c}(x')\Bigr) H^{-\frac{1}{2}} G R(y)\; \varepsilon (x-y),\cr
F^{(21)}(x,y)&= - \Bigr( F^{(12)}(y,x)\Bigr)^{T},\cr
F^{(22)}(x,y)&=\frac{2 \pi}{N}\; \varepsilon (x-y)\cr +\frac{8 \pi}{N}
R(x)^{T} & G H^{-\frac{1}{2}}
P exp \Bigl(\int_{x}^{y} dx' B_{c} m_{c}(x')\Bigr) H^{-\frac{1}{2}}
G R(y)\; \varepsilon (x-y).&(3.7)\cr}
$$
In these equations, the P in front of the exponential means that the
exponential is to be path ordered, and $ R^{T} $ is the transpose of the
matrix R. We have also assumed, as we shall throughout the paper, that
H is a non-singular matrix.

It is now easy to compute any desired PB in terms of F's. Of particular
interest are the PB's of the conserved currents given by
eq.(2.8). After some simple algebra, we find that
$$
\{ J_{-,a}(x), J_{-,b}(y)\}= -\frac{N}{4 \pi}\delta_{ab} \delta^{\prime}
(x-y)+ f_{abc} \delta(x-y) J_{-,c}(x).\eqno(3.8)
$$
As expected, the conserved currents satisfy an affine Lie algebra, and
consequently,
 they are particularly easy to quantize. However, they do not
form a complete set and must be supplemented by additional variables.  We
 found it convenient to work with the complete set $J_{-,a}(x)$ plus
$m_{a}(x)$ of eq.(3.5b). In addition to the PB's given by (3.8), we
 therefore
need the PB's between the J's and m's and between the m's themselves. The
results simplify slightly by redefining
$$
M_{a}(x)=( H^{\frac{1}{2}})_{ab}\: m_{b}(x).\eqno(3.9)
$$
Again, after little bit of algebra, we find that
$$
\{J_{-,a}(x), M_{b}(y)\}= 0,\eqno(3.10a)
$$
and
$$
\eqalignno{
\{M_{a}(x)& ,M_{b}(y)\} =  -\frac{4 \pi}{N} \delta_{ab}
 \delta^{\prime} (x-y)
+ \frac{4 \pi}{N} \delta (x-y) F_{abc} M_{c}(x)
- \frac{2 \pi}{N} \varepsilon (x-y)\cr A_{caa'} &\Bigl( P exp(\int_{x}^{y}
d x' B_{e}( H^{-\frac{1}{2}})_{ee'} M_{e'}(x'))\Bigr)_{a'b'} A_{bdb'}
M_{c}(x) M_{d}(y),&(3.10b)\cr}
$$
where the constants A and F are defined by
$$
\eqalignno{
A_{abc}  =& - 2 H^{-\frac{1}{2}}_{aa'} H^{-\frac{1}{2}}_{bb'}
(H^{-\frac{1}{2}} G)_{cc'} f_{a'b'c'}
 + 4 (H^{-\frac{1}{2}} G)_{aa'}
(H^{-\frac{1}{2}} G)_{bb'} H^{-\frac{1}{2}}_{cc'} f_{a'b'c'},\cr
F_{abc}  =& H^{-\frac{1}{2}}_{aa'} H^{-\frac{1}{2}}_{bb'}
H^{-\frac{1}{2}}_{cc'} f_{a'b'c'}
 - 8 ( H^{-\frac{1}{2}} G)_{aa'}
(H^{-\frac{1}{2}} G)_{bb'} (H^{-\frac{1}{2}} G)_{cc'}
f_{a'b'c'}.&(3.10c)\cr}
$$
The set of equations (3.8) and (3.10), which give the PB's of the complete
set of dynamical variables $J_{-,a}(x)$ and $M_{a}(x)$, provide the natural
starting point for the quantization of the model.
\vskip 9pt
\noindent {\bf 4. Quantization}
\vskip 9pt

The problem of quantization reduces to  quantizing the J's and M's
separetely, as they commute with each other. Since the quantization
of the affine Lie algebra generated by the J's is standard, from now
on  we will focus exclusively on the problem of
quantizing the non-linear algebra (3.10) generated by the M's. This
algebra, in addition to being non-linear, is also non-local, and its
quantization presents a formidable problem. In fact, because of
difficulties with operator ordering, it is not even clear that the final
 result will be unique. The strategy which we shall follow in this paper
is to convert the classical PB's into the corresponding OPE's (operator
product expansion). There is, however, an obstacle to doing this directly
for the M's in the Heisenberg representation: These fields are functions of
$x_{+}$ as well as $x\equiv x_{-}$ and the OPE's, unlike the PB's, will in
general depend on both of these variables. This is the typical problem one
encounters in trying to generalize equal time commutators into OPE's. In
principle, the unknown $x_{+}$ dependence can be determined through the
equations of motion; but instead , we found it easier to go into
the Schroedinger representation by defining a new set of fields which
coincide with the M's at one fixed value of $x_{+}$ and which do not
depend on $x_{+}$. In order not to complicate the notation, we shall use
the same letter M for these fields, and to avoid confusion, it will be
understood that from now on we will be dealing exclusively with the new
M's, which are chiral by definition:
$$
\partial_{+} M_{a}=0.\eqno(4.1)
$$
We now face the problem of converting the PB's of eq.(3.10) into the OPE's
for the new M's. Unable to find a solution in closed form, we resort to
to an expansion in inverse powers of N to make the problem tractable. It is
convenient to define the expansion parameter by
$$
\alpha =(4 \pi/ N)^{\frac{1}{2}},\eqno(4.2a)
$$
and expand M in powers of $\alpha $:
$$
M= \sum_{n=0}^{\infty} \alpha^{n+1} M^{(n)}.\eqno(4.2b)
$$
The PB's of $M^{(n)}$ can be read off from (3.10). For future use, we
exhibit them for values of n up to $n=2$, which is all we need to find
the first order correction to the classical result in the large N
expansion:
$$
\eqalignno{
\{M_{a}^{(0)}(x),M_{b}^{(0)}(y)\}&= -\delta_{ab} \delta^{\prime} (x-y),\cr
\sum_{n=0}^{1} \{M_{a}^{(n)}(x),M_{b}^{(1-n)}(y)\}&
= F_{abc} M_{c}^{(0)}(x) \delta (x-y),\cr
\sum_{n=0}^{2} \{M_{a}^{(n)}(x),M_{b}^{(2-n)}(y)\}& = F_{abc}
M_{c}^{(1)}(x)\delta (x-y)\cr - \frac{1}{2}\varepsilon (x-y) E_{ab,cd}&
M_{c}^{(0)}(x) M_{d}^{(0)}(y),& (4.3a)\cr}
$$
where E is defined by
$$
E_{ab,cd}= A_{cae} A_{bde},\eqno(4.3b)
$$
and A is given by (3.10c). As a check on this result, it is of  interest
to verify that the PB's of (4.3a) satisfy the Jacobi identity. This is
 easily established with the help of the following identity satisfied
 by the constants F and E:
$$
\sum_{cyc.perm.}(F_{a_{1} a_{2} b} F_{b a_{3} c}- E_{a_{2} a_{3}, a_{1} c})
= 0, \eqno(4.3c)
$$
where the sum is over the cyclic permutations of the indices 1, 2 and 3.
This identity follows directly from the Jacobi identity for the structure
constants f.

Our next task is to convert these PB's of the classical fields into the
 OPE's of the quantized fields. There is , of course, some ambiguity in
doing this, which we resolve by adopting the following approach: The OPE of
 the product of two fields is directly obtained from the corresponding PB's
by making the replacements
$$
\eqalignno{
\delta (x-y)& \rightarrow - \frac{1}{2 \pi (x-y)},\cr
\delta' (x-y)& \rightarrow \frac{1}{2 \pi (x-y)^{2}},\cr
\varepsilon (x-y)& \rightarrow \frac{1}{\pi} log(x-y).& (4.4a)\cr}
$$
Doing this in the PB's given by eq.(4.3), we  arrive at the  result, valid
  to second order in $\alpha$, for the OPE's of the product of two fields:
$$
\eqalignno{
M_{a}^{(0)}(x) M_{b}^{(0)}(y)& \cong \frac{1}{2 \pi (x-y)^{2}} \delta_{ab},
\cr
\sum_{n=0}^{1} M_{a}^{(n)}(x) M_{b}^{(1-n)}(y)& \cong - \frac{1}
{4 \pi (x-y)} F_{abc} \Bigl(M_{c}^{(0)}(x)+M_{c}^{(0)}(y)\Bigr),\cr
\sum_{n=0}^{2} M_{a}^{(n)}(x) M_{b}^{(2-n)}(y)& \cong - \frac{1}
{4 \pi (x-y)} F_{abc} \Bigl(M_{c}^{(1)}(x)+M_{c}^{(1)}(y)\Bigr)\cr
- \frac{1}{2 \pi} E_{ab,a'b'}& log(x-y) M_{a'}^{(0)}(x) M_{b'}^{(0)}(y).&
(4.4b)\cr}
$$
 These equations, as well as the other OPE equations we will encounter, are
 understood to include only the terms singular as $x-y \rightarrow 0$.
This is no real loss, since for the applications we have in mind, only the
singular terms are needed. We also note that the right hand
side of these equations have the required symmetry under the interchange
$ x \leftrightarrow y, a \leftrightarrow b$.

The two point OPE's given by the above equation are fundamental in the
 sense that
they can be used as building blocks for constructing the OPE's of the
product of three or more fields to the same order in $\alpha$. We shall carry
out this construction explicitly in the case of product of three and four
fields. The key ingredients of this construction, are the requirements that\\
a)the OPE be symmetric under the permutations of the fields(Bose statistics),\\
b)and that it factorize into the product of two point OPE's for special
configurations of the coordinates. We shall explain this  more fully in
the case of the three point OPE below, but first, we have to adress the
important question of consistency; namely, does this construction the higher
OPE's from the two point OPE lead to a unique answer? The first non-trivial
case to consider is the three point OPE; checking its consistency is
 anologous to verifying
the Jacobi identity in the classical case of the PB's. The three point OPE
in question can be written in several slightly different forms; these are
equivalent up to terms non-singular in the limit of short distances. We
give below one particular form of it:
$$
\eqalignno{
\sum_{n_{1}+n_{2}+n_{3}=n}& M_{a_{1}}^{(n_{1})}(x_{1})
 M_{a_{2}}^{(n_{2})}(x_{2}) M_{a_{3}}^{(n_{3})}(x_{3})
\cong \frac{1}{2 \pi (x_{1} - x_{2})^{2}} \delta_{a_{1},a_{2}}
M_{a_{3}}^{(n)}(x_{3})\cr + Perm. +& W_{a_{1}a_{2}a_{3}}^{(n)}(x_{1},
x_{2},x_{3}), & (4.5)\cr}
$$
where W's are given by
$$
\eqalignno{
W^{(0)}& =  0,\cr
W^{(1)}& =  \frac{1}{ (2 \pi)^{2} (x_{1}- x_{2})(x_{2} - x_{3})
(x_{3}- x_{1})} F_{ a_{1} a_{2} a_{3}},\cr
W^{(2)}& = - \frac{1}{4 \pi^{2}} log(x_{1}- x_{2}) E_{a_{1} a_{2}, b_{1}
b_{2}} \Bigl(\frac{\delta_{b_{1} b_{2}}}{ (x_{1}- x_{2})^{2}}
M_{a_{3}}^{(0)}(x_{3}) + \frac{\delta_{b_{1} a_{3}}}{(x_{1}- x_{3})^{2}}
M_{b_{2}}^{(0)}(x_{2})\cr
 +& \frac{\delta_{b_{2} a_{3}}}{(x_{2}- x_{3})^{2}}
M_{b_{1}}^{(0)}(x_{1})\Bigr)
+ Perm. + Z_{a_{1} a_{2} a_{3}}(x_{1}, x_{2}, x_{3}),& (4.6a)\cr}
$$
and  Z by
$$
\eqalignno{
Z & = \frac{1}{(2 \pi)^{2} (x_{3}-x_{1})(x_{1}- x_{2})}(F_{a_{3} a_{1} b}
F_{b a_{2} c} - E_{a_{1} a_{2},a_{3} c}) M_{c}^{(0)}(x_{2})\cr
+ & \frac{1}{(2 \pi)^{2} (x_{3}- x_{2})(x_{1}-x_{2})}(F_{a_{2} a_{3} b}
F_{b a_{1} c} - E_{a_{3} a_{1}, a_{2} c}) M_{c}^{(0)}(x_{1})\cr
+ & \frac{1}{8 \pi^{2}(x_{3} - x_{1})} F_{a_{3} a_{1} b} F_{b a_{2} c}
M_{c}^{\prime (0)}(x_{2})
 + \frac{1}{8 \pi^{2}(x_{2} -x_{3})}
F_{a_{2} a_{3} b}  F_{b a_{1} c} M_{c}^{\prime (0)}(x_{1})\cr
+& \frac{1}{8 \pi^{2}(x_{1}- x_{2})}( - F_{a_{1} a_{2} b} F_{b a_{3} c}
+ E_{a_{2} a_{3}, a_{1} c}) M_{c}^{\prime (0)}(x_{3}).& (4.6b)\cr}
$$
In these equations, the symbol ``Perm.'' means that the main term has to
be symmetrized in the indices 1, 2 and 3 by adding all of the necessary
permutations in these indices to it. As a result, except for Z, all the
 terms on the right hand side of these equations are manifestly symmetric
and unique. On the other hand, the term Z is not symmetric and therefore
not unique: One could permute its indices and obtain a different version
of Z. These different versions, however, differ only by terms non-singular
in the short distance limit, and they are therefore equivalent as explained
above. This can easily be shown with the help of
(4.3c), the same identity that was needed to establish the Jacobi identity
for the PB's.

It remains to verify factorization, which is defined as follows: The
OPE given by eq.(4.6) is valid in the limit when the differences between
$x_{1}$, $x_{2}$ and $x_{3}$ all go to zero, with no further restrictions.
Now let us consider the limit
$$
x_{1}- x_{3}\rightarrow 0,\: x_{1,3}- x_{2}\rightarrow 0 ,\eqno(4.7a)
$$
subject, however, to the restriction
$$
|x_{1}- x_{3}| \ll |x_{1}- x_{2}|. \eqno(4.7b)
$$
In this limit, the three point OPE can then be constructed by the double
application of the two point OPE's( eq.4.4b): 1 and 3 are first combined
into an OPE, and the result is then combined with 2. This corresponds to
 taking the limit $x_{1}- x_{3}\rightarrow 0$ first, before letting
 $x_{1}- x_{2}\rightarrow 0$, as in eq.(4.7). This should then agree with
the three point OPE(eq.4.6), in the same limit given  by (4.7b). This means
that in making the comparison, we are allowed first to drop terms non-
singular in $x_{1}-x_{3}$, and then terms non-singular in the differences
$x_{1}- x_{2}$ and $x_{3}- x_{2}$.It is then  fairly
easy to verify that the three point OPE given by (4.5) and (4.6) does
satisfy factorization. As an example, we exhibit below the particular term
corresponding to Z  in the factorized form of the three point OPE:
$$
\eqalignno{
\Bigl(M_{a_{1}} (x_{1})M_{a_{3}}(x_{3})\Bigr)& M_{a_{2}}(x_{2})\cr
\cong \frac{1}{16 \pi^{2}(x_{3}- x_{1})}& F_{a_{3} a_{1} b} F_{b a_{2} c}
\Bigl(\frac{ M_{c}^{(0)}(x_{1})+ M_{c}^{(0)}(x_{2})}{x_{1}-x_{2}}+
\frac{M_{c}^{(0)}(x_{3}+ M_{c}^{(0)}(x_{2})}{x_{3}- x_{2}}\Bigr)\cr
\cong \frac{1}{4 \pi^{2}(x_{3}- x_{1})}& F_{a_{3} a_{1} b}F_{b a_{2} c}
 \Bigl(\frac{1}{x_{1}- x_{2}} M_{c}^{(0)}(x_{2})+ \frac{1}{2}
 M_{c}^{\prime (0)}(x_{2})\Bigr).&(4.8)\cr}
$$
 Up to non-singular terms as explained above,
this agrees with Z given by eq.(4.6b), and one can then check various other
 terms in a similar fashion. There are, of course, in addition to
(4.7), two other order of limits to consider. These are related to (4.7)
by permutations, and in view of the symmetry of the three point OPE,
factorization in these channels follows from factorization in the channel
we have already considered. We have therefore shown that the two point
OPE's given by eq.(4.3) satisfy the consistency condition analogous to the
Jacobi identity in the case of the PB's. Here, consistency means that
 conditions a) and b) lead to an essentially unique result. Our proof holds
 only to second order in $\alpha$, however, apart from increasing
algebraic complexity, we see no fundamental obstacle to extending this
argument to higher orders. There still remains the question of establishing
the consistency of the higher point OPE's. In the classical case, the Jacobi
identity for the double PB, which is the anologue of the three point OPE,
is all that is needed; the consistency conditions for the higher multi
PB's follow from it. We believe the same to be true for the OPE's: The
consistency of the three point OPE guarantees the same for the higher point
 OPE's. Although we have no general proof of this statement, we have
established it for the four point OPE, which will be studied in the next
section.
\vskip 9pt
\noindent {\bf 5. The Conformal Algebra}
\vskip 9pt

The action given by eq.(2.5) is  invariant under classical conformal
transformations generated by the stress tensor. It can easily be shown that
the PB's of the stress tensor satisfy the classical Virasoro algebra,
without, of course, the central term. Quantum mechanically, the OPE's of
the product of the stress tensor with itself, apart from the central charge,
 will have anomalous terms
signaling the breakdown of conformal invariance. In our approach, conformal
invariance is restored by imposing conditions on the coupling constants to
get rid of the anomalous terms and to reestablish the Virasoro algebra.
We start with the classical expression for the $(-,-)$ component of the stress
tensor T, derived from the bosonic part of the action given by (2.5). In
out light cone approach, with $x_{+}$ identified with time, this is the
only component we need to consider. Defining
$$
T_{-,-}= T + \tilde{T},\eqno(5.1a)
$$
we have,
$$
\eqalignno{
T=& \frac{N}{4 \pi} M_{a} M_{a},\cr
\tilde{T}=& \frac{4 \pi}{N} J_{-,a} J_{-,a}.& (5.1b)\cr}
$$

One can easily show that
the PB of T with $\tilde{T}$ vanishes and each one seperately satisfies
a classical Virasoro algebra. In the case of $\tilde{T}$, when the classical
 $J_{-}$'s are replaced by
 their quantum version, one is faced by the problem of defining the product
of two J's at the same point. This problem is easily resolved by first
splitting the two coordinates, subtracting the short distance singularity,
and then taking the equal coordinate limit. This results in the standard
Sugawara construction, with N shifted as in (2.6a). We will use exactly
the same procedure in defining the quantum expression for T by letting
$$
T(x)=\frac{\pi}{\alpha^{2}} \lim_{x- y \rightarrow 0}(M_{a}(x)M_{a}(y)-
 sing.terms).\eqno(5.2)
$$
The singular terms to be subtracted before taking the limit $x- y\rightarrow
0 $ are given by the right hand side of eq.(4.4b). This prescription therefore
guarantees a finite result for T.
Starting with this definition of T, we are now in a position to check whether
the Virasoro algebra, with the central charge c,
$$
T(x)T(y)\cong \frac{1}{(x- y)^{2}}(T(x)+ T(y)) +\frac{c}{2 (x- y)^{4}},
\eqno(5.3)
$$
is satisfied. This will be done by first constructing the four point OPE,
and then converting it to the OPE of two T's by taking the limits indicated
in (5.2). Again, the calculation is carried out to second order in $\alpha$.
The four point OPE can uniquely be determined from factorization and symmetry
in a manner identical to the construction of the three point OPE. We will
therefore skip the details and present the term  second order in
 $\alpha$, which is the only non-trivial term needed to verify eq.(5.3):
$$
\eqalignno{
\sum_{n_{1}+ n_{2}+n_{3}+ n_{4}=2}& M_{a_{1}}^{(n_{1})}(x_{1})M_{a_{2}}^
{(n_{2})}(x_{2})M_{a_{3}}^{(n_{3})}(x_{3})M_{a_{4}}^{(n_{4})}(x_{4})\cr
\cong \frac{\delta_{a_{1}a_{2}}}{2 \pi (x_{1}- x_{2})^{2}}& M_{a_{3}}^{(1)}
(x_{3})M_{a_{4}}^{(1)}(x_{4})+ Perm.+\frac{1}{16 \pi^{2}(x_{1}- x_{2})
(x_{3}- x_{4})}\cr
F_{a_{1}a_{2}b}F_{a_{3}a_{4}c}& (M_{b}^{(0)}(x_{1})+M_{b}^{(0)}(x_{2}))
(M_{c}^{(0)}(x_{3})+M_{c}^{(0)}(x_{4}))+ Perm.\cr
+M^{(0)}_{a_{4}}(x_{4})& Z_{a_{1}a_{2}a_{3}}(x_{1},x_{2},x_{3})+ Perm.
-\frac{1}{4 \pi^{2}}log(x_{1}-x_{2})E_{a_{1}a_{2},b_{1}b_{2}}\cr
\Bigl(\frac{\delta_{b_{1}b_{2}}}{(x_{1}- x_{2})^{2}}& M_{a_{3}}^{(0)}
M_{a_{4}}^{(0)}(x_{4})+\frac{\delta_{b_{1}a_{3}}}{(x_{1}- x_{3})^{2}}
M_{b_{2}}^{(0)}(x_{2})M_{a_{4}}^{(0)}(x_{4})\cr
+\frac{\delta_{b_{2}a_{3}}}{(x_{2}- x_{3})^{2}}& M_{b_{1}}^{(0)}(x_{1})
M_{a_{4}}^{(0)}(x_{4})+ \frac{\delta_{b_{1}a_{4}}}{(x_{1}- x_{4})^{2}}
M_{b_{2}}^{(0)}(x_{2})M_{a_{3}}^{(0)}(x_{3})\cr
+\frac{\delta_{b_{2}a_{4}}}{(x_{2}- x_{4})^{2}}& M_{b_{1}}^{(0)}(x_{1})
M_{a_{3}}^{(0)}(x_{3})+ \frac{\delta_{a_{3}a_{4}}}{(x_{3}- x_{4})^{2}}
M_{b_{1}}^{(0)}(x_{1})M_{b_{2}}^{(0)}(x_{2})\Bigr)+ Perm.\cr
+\Delta_{a_{1}a_{2}a_{3}a_{4}} & (x_{1},x_{2},x_{3},x_{4}).&(5.4a)\cr}
$$
In this equation, ``Perm.'' again stands for the terms to be added to
completely symmetrize  the right hand side, and Z is given by (4.6b).
Notice that, except for $\Delta$, all of the terms given above are
quadratic in the field M and therefore contribute only to the first term
 on the right hand side of eq.(5.3). On the other hand,
$\Delta$ is a c-number term that contributes only to the central charge; it
is given by
$$
\eqalignno{
\Delta =\frac{A_{a_{1}a_{2}a_{3}a_{4}}}{(x_{1}- x_{2})(x_{3}- x_{4})}&
\Bigl(\frac{1}{(x_{1}- x_{3})(x_{2}- x_{4})}+ \frac{1}{(x_{2}- x_{3})
(x_{1}- x_{4})}\Bigr)\cr
+\frac{A_{a_{1}a_{3}a_{2}a_{4}}}{(x_{1}- x_{3})(x_{2}- x_{4})}&\Bigl(
\frac{1}{(x_{1}- x_{2})(x_{3}- x_{4})}+ \frac{1}{(x_{1}- x_{4})
(x_{3}- x_{2})}\Bigr)\cr
+ \frac{A_{a_{1}a_{4}a_{2}a_{3}}}{(x_{1}- x_{4})(x_{2}- x_{3})}&
\Bigl(\frac{1}{(x_{1}- x_{2})(x_{4}- x_{3})}+ \frac{1}{(x_{1}- x_{3})
(x_{4}- x_{2})}\Bigr)\cr
- \frac{1}{(2 \pi)^{3}} log(x_{1}- x_{2})& E_{a_{1}a_{2},b_{1}b_{2}}\Bigl(
\frac{\delta_{b_{1}b_{2}}\delta_{a_{3}a_{4}}}{(x_{1}- x_{2})^{2}(x_{3}-
x_{4})^{2}}\cr
+\frac{\delta_{b_{1}a_{4}}\delta_{b_{2}a_{3}}}{(x_{1}- x_{4})^{2}(x_{2}-
x_{3})^{2}}& + \frac{\delta_{b_{1}a_{3}}\delta_{b_{2}a_{4}}}{(x_{1}-
x_{3})^{2}(x_{2}- x_{4})^{2}}\Bigr)+ Perm.,&(5.4b)\cr}
$$
where,
$$
A_{a_{1}a_{2}a_{3}a_{4}}= \frac{1}{24 \pi^{3}}(F_{a_{1}a_{2}b}F_{a_{3}a_{4}
b}+ E_{a_{1}a_{3},a_{2}a_{4}}).\eqno(5.4c)
$$
Given the four point OPE, one can then construct the OPE for the product
of two T's by carrying out the limit indicated by (5.2). To the given order
in $\alpha$, the result is
$$
\eqalignno{
T(x)T(y)\cong & \frac{\pi}{\pi (x-y)^{2}}\Bigl(M_{a}^{(0)}(x)
M_{a}^{(0)}(x)+ M_{a}^{(0)}(y)M_{a}^{(0)}(y)\cr
+\alpha^{2} M_{a}^{(1)}(x)M_{a}^{(1)}(x)& + \alpha^{2} M_{a}^{(1)}(y)
M_{a}^{(1)}(y)\Bigr)
-\frac{\alpha^{2}}{4 (x-y)^{2}} (F_{aa'b}F_{aa'c}+2 E_{aa,bc})\cr
\Bigl(M_{b}^{(0)}(x)M_{c}^{(0)}(x)& + M_{b}^{(0)}(y)M_{c}^{(0)}(y)\Bigr)
+\frac{c}{2(x-y)^{4}},&(5.5a)\cr}
$$
where, to the same order, the central charge c is given by
$$
c= D - \frac{\alpha^{2}}{4 \pi}(F_{abc}F_{abc}+ E_{aa,bb}),\eqno(5.5b)
$$
and D is the dimension of the flavor algebra, equal to $n^{2}-1$ in case of
$SU(n)$. Clearly, to satisfy the conformal algebra(eq.(5.3)),  the condition
$$
F_{aa'b} F_{aa'c}+ 2 E_{aa,bc}= 0.\eqno(5.6a)
$$
has to be imposed. With the help of eqns.(3.10c) and 4.3b), this can be
written as the following condition on the coupling constants $G_{ab}$:
$$
\eqalignno{
Tr&\Bigl((1-4 G^{2})^{-1}f_{a}(1-4 G^{2})^{-1}f_{b}\Bigr)
 -2 Tr\Bigl(f_{a}f_{b}(1-4G^{2})^{-1}\Bigr)\cr
-8& G_{bb'} Tr\Bigl(G (1-4G^{2})^{-1}f_{a}G (1-4G^{2})^{-1}f_{b'}\Bigr)\cr
-8& G_{aa'} Tr\Bigl(G (1-4G^{2})^{-1}f_{a'}G (1-4G^{2})^{-1}f_{b}\Bigr)\cr
 + 4& G_{aa'}G_{bb'}\Bigl(Tr\Bigl((1-4G^{2})^{-1}f_{a'}
(1-4G^{2})^{-1}f_{b'}\Bigr)- Tr(f_{a'}f_{b'})\Bigr)=0,&(5.6b)\cr}
$$
where we have used the matrix notation for both G and the
structure constants f, with the definition
$$
f_{a}\rightarrow (f_{a})_{bc}\equiv f_{abc}.\eqno(5.6c)
$$

Eq.(5.6b), which determines the conformal points in the coupling constant
space, is the main result of this paper. We have investigated this equation
in the case of $SU(2)$ and found that, excluding the cases when H becomes
singular, it has it has two types of solutions.
In both solutions, G is diagonal:
$$
G_{ab}=g_{a}\delta_{ab},\eqno(5.7a)
$$
and the first type of solution is $SU(N)$ symmetric, with G proportional to
the unit matrix. The constant of proportionality is restricted to two
possible values:
$$
g_{a}=g= \frac{\sqrt{2}+1}{4}\pm \frac{1}{4}\sqrt{2\sqrt{2}-1}.\eqno(5.6b)
$$
These probably correspond to some of the Dashen-Frishman[10] conformal
points. In the other type of solution, one can take two eigenvalues, say
$g_{1}$ and $g_{2}$ equal and $g_{3}$ to have the opposite sign:
$$
g_{1}=g_{2}=-g_{3}=-g,\eqno(5.6b)
$$
where g is given by (5.6b). This solution has only $U(1)$ symmetry and its
significance is less clear. Although we have not investigated higher flavor
groups in detail, it is likely that they also admit solutions.
\vskip 9pt
\noindent{\bf 6. Conclusions}
\vskip 9pt

The main result of this paper are the conditions for conformal invariance
given by eqns.(5.6). The coupling constants that satisfy these equations
should then correspond to conformal fixed points invariant generalized
 Thirring model, at least to the given order in the large N expansion.
The road should therefore be open for the application of these results
to the construction new string models.
 However, several problems remain to be solved. For example, it would be
 nice to work out the higher order corrections in $\alpha$ to the OPE of
 eq.(4.4b); or better yet, to discover the exact expression. It is also of
 some interest to be able to compare the
Hamiltonian approach of this paper with Lagrangian approach of [14].
Another interesting suggestion [17] to verify or disprove is the idea that
the stress tensor of the generalized Thirring model at the conformal points
admits of an affine Virasoro construction [7] in terms of chiral currents
that satisfy an affine algebra. In our case,  eq.(4.4b) does not represent
 an affine algebra, so the question is whether through a non-linear and
possibly non-local transformation of the field M, it is nevertheless possibe
 to map (4.4b) into an affine algebra. We have not succeeded in doing so;
however, the question still remains open.
\vskip 9pt
\noindent{\bf Acknowledgements}
\vskip 9pt

I would like to thank Martin Halpern for an interesting conversation. I would
also like to thank Andre Neveu for hospitality at the University of
Montpellier, where part of this work was done.
\newpage
{\bf References}
\begin{enumerate}
\item A.Belavin, A.M.Polyakov and A.B.Zamolodchikov, Nuc. Phys. {\bf B241}
(1984) 333.
\item M.Green, J.H.Schwartz and E.Witten, ``{\em Superstring Theory}'',
Cambridge University Press (1987).
\item E.Witten, Commun. Math. Phys. {\bf 92} (1984) 455.
\item E.Witten, Nuc. Phys. {bf B223} (1983) 422; I.Antoniadis and C.Bachas,
Nuc. Phys. {\bf B278} (1986) 343.
\item P.Candelas, G.Horowitz, A.Strominger and E.Witten, Nuc. Phys.
{\bf 258} (1985) 46.
\item K.Bardakci and M.B.Halpern, Phys. Rev. {\bf D3} (1971) 2493.
\item M.B.Halpern and E.Kiritsis, Mod. Phys. Lett. {\bf A4} (1989) 1373;
Erratum ibid. {\bf A4} (1989) 1797. For a Lagrangian description, see
M.B.Halpern and J.P.Yamron, Nuc. Phys. {\bf B351} (1991) 333.
\item G.'t Hooft, Nuc. Phys. {\bf B75} (1974) 461.
\item K.Bardakci, Nuc. Phys. {\bf B401} (1993) 168.
\item R.Dashen and Y.Frishman, Phys. Rev. {\bf D11} (1975) 2781.
\item C.Destri, Phys. Lett. {\bf B210} (1988) 173.
\item D.Kutasov, Phys. Lett. {\bf B227} (1989) 68.
\item D.Karabali, Q.H.Park and H.J.Schnitzer, Nuc. Phys. {\bf B323} (1989)
572, Phys. Lett. {\bf B205} (1988) 267.
\item A.A.Tseytlin, ``{\em On a Universal Class of WZW-Type Conformal
Models}'',
CERN-TH.7068/93, hep-th/9311062.
\item C.G.Callan, D.Friedan, E.Martinec and M.J.Perry, Nuc. Phys. {\bf B262}
(1985) 593.
\item A.M.Polyakov and P.B.Wiegmann, Phys. Lett. {\bf B131} (1983) 121,
{\bf B141} (1984) 223.
\item K.Bardakci, M.Crescimanno and E.Rabinovici, Nuc. Phys. {\bf B344}
(1990) 344.
\item O.A.Soloviev, ``{\em Conformal Non-Abelian Thirring Models}'',
QMW 93-19, hep-th/9307163.
\end{enumerate}

\end{document}